\documentclass[sigconf]{acmart}

\usepackage{multirow}
\usepackage{listings}
\usepackage{subcaption}
\usepackage[T1]{fontenc}
\usepackage{fancybox,framed}
\usepackage{xcolor}
\usepackage{graphicx}
\usepackage{longtable}
\usepackage{lscape}
\usepackage{todonotes}
\usepackage{multirow}
\usepackage{colortbl}
\definecolor{Gray}{gray}{0.9}
\usepackage[flushleft]{threeparttable}
\usepackage{listings}
\usepackage{enumitem}
\usepackage{tablefootnote}
\usepackage{float}
\usepackage{booktabs}
\usepackage{tcolorbox}
\usepackage{tabularx}
\usepackage{enumitem}
\usepackage{mdframed}

\tcbuselibrary{most}
\rowcolors{2}{gray!10}{white}  % Start with row 2, alternate between light gray and white

\AtBeginDocument{%
  }

\copyrightyear{2025}
\acmYear{2025}
\setcopyright{cc}
\setcctype{by}
\acmConference[ICER 2025 Vol. 1]{ACM Conference on International Computing
Education Research V.1}{August 3--6, 2025}{Charlottesville, VA, USA}
\acmBooktitle{ACM Conference on International Computing Education Research
V.1 (ICER 2025 Vol. 1), August 3--6, 2025, Charlottesville, VA, USA}
\acmDOI{10.1145/3702652.3744219}
\acmISBN{979-8-4007-1340-8/2025/08}
\begin{document}

%%
%% The "title" command has an optional parameter,
%% allowing the author to define a "short title" to be used in page headers.
\title[The Effects of GitHub Copilot on Computing Students' Programming Effectiveness, Efficiency, and Processes]{The Effects of GitHub Copilot on Computing Students' Programming Effectiveness, Efficiency, and Processes in Brownfield Programming Tasks}

%%
%% The "author" command and its associated commands are used to define
%% the authors and their affiliations.
%% Of note is the shared affiliation of the first two authors, and the
%% "authornote" and "authornotemark" commands
%% used to denote shared contribution to the research.
\author{Md Istiak Hossain Shihab, Christopher Hundhausen, Ahsun Tariq}
\author{Summit Haque, Yunhan Qiao, Brian Mulanda}

\affiliation{%
\department{Software Engineering Education Assessment and Analytics Lab} 
\department{School of Electrical Engineering and Computer Science}
 \institution{Oregon State University}
 \city{Corvallis}
 \state{Oregon}
 \country{USA}}

 \email{{shihabm, chris.hundhausen, tariqa, haquesu, qiaoy, mulandab}@oregonstate.edu}

\renewcommand{\shortauthors}{Shihab et al.}

%%
%% The abstract is a short summary of the work to be presented in the
%% article.
\begin{abstract}
When graduates of computing degree programs enter the software industry, they will most likely join teams working on legacy code bases developed by people other than themselves. In these so-called \textit{brownfield} software development settings, generative artificial intelligence (GenAI) coding assistants like GitHub Copilot are rapidly transforming software development practices, yet the impact of GenAI on student programmers performing brownfield development tasks remains underexplored. This paper investigates how GitHub Copilot influences undergraduate students' programming performance, behaviors, and understanding when completing \textit{brownfield programming tasks} in which they add new code to an unfamiliar code base. We conducted a controlled experiment in which 10 undergraduate computer science students completed highly similar brownfield development tasks with and without Copilot in a legacy web application. Using a mixed-methods approach combining performance analysis, behavioral analysis, and exit interviews, we found that students completed tasks 35\% faster (\textit{p} < 0.05) and made 50\% more solution progress (\textit{p} < 0.05) when using Copilot. Moreover, our analysis revealed that, when using Copilot, students spent 11\% less time manually writing code (p < 0.05), and 12\% less time conducting web searches (\textit{p} < 0.05), providing evidence of a fundamental shift in how they engaged in programming. 
%When using Copilot, higher-performing students tended to be more selective in their use of AI-generated code, preferring granular inline suggestions over adoption of code blocks wholesale.
In exit interviews, students reported concerns about not understanding how or why Copilot suggestions work.
%highlighting a crucial tension: GenAI may promote brownfield programming efficiency at the cost of diminished learning.
This research suggests the need for computing educators to develop new pedagogical approaches that leverage GenAI assistants' benefits while fostering reflection on how and why GenAI suggestions address brownfield programming tasks. Complete study results and analysis are presented at \href{https://ghcopilot-icer.github.io/}{ghcopilot-icer.github.io}.
\end{abstract}

%%
%% The code below is generated by the tool at http://dl.acm.org/ccs.cfm.
%% Please copy and paste the code instead of the example below.
%%


%%
%% Keywords. The author(s) should pick words that accurately describe
%% the work being presented. Separate the keywords with commas.

%% A "teaser" image appears between the author and affiliation
%% information and the body of the document, and typically spans the
%% page.
% \begin{teaserfigure}
%   \includegraphics[width=\textwidth]{sampleteaser}
%   \caption{Seattle Mariners at Spring Training, 2010.}
%   \Description{Enjoying the baseball game from the third-base
%   seats. Ichiro Suzuki preparing to bat.}
%   \label{fig:teaser}
% \end{teaserfigure}

% \received{20 February 2007}
% \received[revised]{12 March 2009}
% \received[accepted]{5 June 2009}

\newenvironment{changemargin}[2]{%
\begin{list}{}{%
\setlength{\topsep}{6pt}%
\setlength{\leftmargin}{#1}%
\setlength{\rightmargin}{#2}%
\setlength{\parsep}{\parskip}%
}%
\item[]}{\end{list}}

\newcommand{\researchquestion}[1]{%
  \begin{list}{}{%
    \setlength{\topsep}{6pt}%
    \setlength{\leftmargin}{0.25in}%
    \setlength{\rightmargin}{0.25in}%
    \setlength{\parsep}{\parskip}%
  }
  \item[]
    \doublebox{%
      \begin{minipage}{4.75in}
        \em #1
      \end{minipage}%
    }
  \end{list}%
}
%%
%% This command processes the author and affiliation and title
%% information and builds the first part of the formatted document.
\maketitle

\setcounter{footnote}{0}
\footnotetext[\value{footnote}]{© Author 2025. This is the author's version of the work. It is posted here for your personal use. Not for redistribution. The definitive Version of Record will be published in Proceedings of the 2025 ACM Conference on International Computing Education Research (ICER '25).}
\addtocounter{footnote}{-1}

\section{Introduction}
\label{sections-s1-introduction}
Contemporary software engineering practices are being revolutionized by the emergence of generative artificial intelligence-based (GenAI) programming assistance tools ~\cite{sauvola2024future, alshammari2022trends}. GitHub Copilot~\cite{copilot}, which uses large language models to generate contextual code suggestions and code explanations, represents a particularly important tool in this domain, given its strong integration into the Visual Studio Code integrated development environment used by 73.6\% of software developers~\cite{stackoverflowsurvey}.

Although recent studies demonstrate the potential of GenAI tools to enhance professional developer productivity ~\cite{peng2023impact, klemmer2024using} and novice programming~\cite{gardella2024performance, peng2023impact}, we know little about how GenAI assistants influence students when performing  \textit{brownfield programming tasks}---that is, tasks that involve enhancing unfamiliar code bases written by others.  
Brownfield programming tasks present unique challenges for student programmers ~\cite{feathers2004Working}. Unlike the \textit{greenfield} programming tasks commonly assigned in academic settings, where students start from scratch, brownfield programming tasks require students to integrate new code into existing code bases, which, because they are written by others, can be difficult to understand and modify ~\cite{srinivas2016analysis}. 

The use of GenAI coding assistants in this context raises important research questions. Of particular interest is how student programmers, who are still developing software engineering expertise, interact with, evaluate, and ultimately integrate GenAI-generated code suggestions into unfamiliar code bases. Indeed, while GenAI assistants can provide immediate coding assistance, they may not always generate suggestions that align with the coding patterns and practices in an existing code base ~\cite{dakhel2023github}. Understanding how students make decisions about these suggestions and integrate them into legacy code is crucial for developing appropriate strategies and guidelines for teaching the brownfield software development skills that students will ultimately need for jobs in the software profession.

Motivated by the need to understand how GenAI assistance influences students' programming outcomes and processes in brownfield development tasks, we present an experimental study to address the following overarching research question:

% Our study focuses on understanding the impact of GitHub Copilot on student programmers working with legacy code bases through a controlled experiment. 
% We examine both quantitative metrics of performance and qualitative aspects of student decision-making processes. 
% This research is particularly timely as educational institutions and industry partners grapple with ~\cite{wang2023exploring, wang2023towards, schon2023ai} how to effectively incorporate AI assistants into software engineering education while ensuring students develop robust programming skills and understanding.

% \subsection{Research Problem and Questions}

% The increasing adoption of AI coding assistants like GitHub Copilot raises important questions about their impact on software development, particularly for student programmers working with unfamiliar legacy code bases. While these tools promise to enhance productivity through automated code suggestions, their effectiveness and influence on students' development practices in real-world scenarios remains unclear. This motivates the overarching research question addressed by this study:

% \noindent\textbf{Primary Research Question:}
% \begin{quote}
%     \itshape How does Copilot influence student programmers' effectiveness when performing brownfield programming tasks?
% \end{quote}

\vspace{2mm}

\begin{mdframed}[linewidth=0.5pt]
\noindent\begin{minipage}[t]{0.20\textwidth}
\textbf{Primary \\Research\\Question}
\end{minipage}%
\hspace{2mm}%
\begin{minipage}[t]{0.75\textwidth}
\textit{How does Copilot influence student programmers' effectiveness when performing brownfield programming tasks?}
\end{minipage}
\end{mdframed}

\vspace{2mm}

\noindent To systematically investigate this question, we break it down into three specific research questions:

\begin{description}
\item[\textbf{RQ1:}] \textit{How does Copilot influence students' performance in brownfield programming tasks?}

\item[\textbf{RQ2:}]\textit{How does Copilot influence students' programming processes when working on brownfield programming tasks?}

\item[\textbf{RQ3:}] \textit{How do students' usage patterns with Copilot differ?}
\end{description}

To address these questions, we conducted a within-subjects experiment with two conditions: \textsc{No Copilot} and \textsc{Copilot}. Our results showed that (a) students completed brownfield coding tasks significantly more efficiently with Copilot; (b) students' programming processes differed markedly with and without Copilot; (c) students who adopted Copilot-generated code at a more granular level tended to be more successful than students who adopted Copilot suggestions wholesale; and (d) students expressed concern over their lack of understanding of how and why Copilot's suggestions worked. Our mixed-methods study contributes the first-ever empirical account of the impact of GenAI on students' effectiveness, accuracy, processes and perceptions when performing brownfield programming tasks.

\vspace{-2mm}

\label{sections-s2-related-work}
\section{Related Work}

\subsection{Large Language Models in Software Development}

The advent of GenAI is significantly changing software development practices~\cite{Liang2023UnderstandingTU, tabarsi2025llms}. Sophisticated GenAI coding assistants such as GitHub Copilot~\cite{copilot}, ChatGPT~\cite{welsby2023chatgpt}, and Gemini~\cite{team2023gemini} are becoming tightly integrated into contemporary integrated development environments. Through such integrations, these tools not only support context-aware conversations, but also provide auto-complete suggestions and contextualized code explanations. The ability to engage with GenAI to explore design alternatives, generate tailored code solutions, and obtain contextualized explanations of code is transforming modern software development into a collaborative process involving humans and GenAI.

%Despite their promise, empirical studies have shown that the effectiveness of GenAI tools varies markedly across different programming domains and complexity levels~\cite{siam2024programming}. Particularly striking is the emergence of advanced models like GPT-4, which—when leveraging optimized prompt strategies—can surpass 85\% of human competitors in coding competitions~\cite{hou2025comparing}. Such remarkable capabilities underscore how large language models are fundamentally reshaping the entire software development lifecycle, from initial conceptualization through implementation to ongoing maintenance, suggesting a paradigm shift in how programming tasks are approached and executed.

% \vspace{-2mm}
\subsection{GenAI Coding Assistants: Performance and Adoption}

GenAI coding assistants have demonstrated considerable potential to enhance developer productivity. Empirical studies have quantified these productivity gains, with one study documenting average productivity increases of 24\% among developers using GitHub Copilot, with junior developers enjoying 44\% productivity improvements~\cite{ng2024harnessing}. Another study have found that GitHub Copilot improved developer efficiency by 55.8\%~\cite{peng2023impact} in a greenfield programming task. Beyond speed enhancements, these tools have shown capabilities in automating routine tasks like code completion, documentation generation, and test case creation~\cite{sergeyuk2025using}.

The usage patterns of these tools vary significantly across different software development activities. O'Brien et al.~\cite{o2025scientists} conducted a comprehensive investigation into how scientists across disciplines leverage large language models for programming support in their research, identifying distinct usage patterns and verification strategies employed by early adopters. Their findings highlighted that professionals often used these tools as information retrieval mechanisms for navigating unfamiliar programming languages and libraries.

A large-scale survey focusing on specific software development activities revealed that developers find writing tests and natural-language artifacts to be the least enjoyable activities and most frequently delegate these tasks to GenAI assistants~\cite{sergeyuk2025using}. The study identified several obstacles to wider adoption, including lack of project-size context awareness and organizational policies restricting AI tool usage in certain environments.

Despite their capabilities, these tools have signficant limitations. Research on GitHub Copilot has found that professional programmers are often reluctant to use these assistants because they are not capable of specifically addressing functional or non-functional requirements, and controlling the tool can be difficult~\cite{liang2024large}. Since generated code often contains unfamiliar APIs, understanding which part of the input is causing faulty outputted code is often difficult, which has forced people to give up on outputted code entirely. These findings underscore the importance of maintaining oversight when incorporating GenAI-generated code into production systems.

The ethical dimensions of AI-assisted programming have also received attention. Research has emphasized that while GenAI can drastically improve the efficiency of software development, it should be viewed as a complementary tool rather than a replacement for traditional programming skills~\cite{atemkeng2024ethics}. This perspective aligns with broader concerns about maintaining programmer agency and understanding in increasingly GenAI-augmented development environments.
% \vspace{-2mm}
\subsection{GenAI Coding Assistants in Computer Science Education}

Undergraduate computing education has been disrupted by the emergence of GenAI technologies, which are causing computer science degree programs to fundamentally reexamine their pedagogical frameworks and assessment methodologies~\cite{bozkurt2023generative, deriba2024computer}. %This educational paradigm shift became more accessible in September 2022 when the GitHub Education program facilitated access to GitHub Copilot for both educators and students, democratizing advanced AI programming assistance within academic environments~\cite{Hollowed_2022}.
Early research by Denny and colleagues demonstrated that even in its initial iteration, GitHub Copilot's capabilities placed it among the top 25\% of students in introductory programming assignments, raising substantial concerns about how educational institutions should respond to GenAI-assisted programming~\cite{finnie2022robots, yu2023reflection}. Building upon these initial findings, a corpus of subsequent investigations has methodically examined the multifaceted impact of GenAI tools on novice programmers' cognitive processes and educational outcomes~\cite{skripchuk2024investigation, scholl2024novice, hashmi2024generative}. A key finding of these investigations is that GenAI tools have varying impacts on novice programmers based on their learning styles, prior AI experience, and academic level, with students showing diverse usage patterns ranging from uncritical acceptance to critical engagement with AI solutions~\cite{scholl2024novice}. 

In a study of the Gemini GenAI assistant in Google Colab environments, Llerena-Izquierdo et al.~\cite{llerena2024innovations} found that an overwhelming majority (91\%) of participants reported that the AI assistance surpassed their educational support expectations. In a related study, Tanay ~\cite{tanay2024exploratory} found that learners predominantly employ GenAI for programming and composition assistance, while perceiving these tools as facilitators of expedited research processes and accelerated project development. However, this study also noted student concerns that GenAI assistants might hurt their ability to retain novel concepts and could be ineffective without a base-level understanding of topics. Pirzado et. al. ~\cite{pirzado2024navigating} found that novice students may find GenAI assistants challenging to use in regular academic activities, but may benefit from them at the beginning of their journey. Tankelevitch and Kewenig, et al. has highlighted the metacognitive demands as a challenge that GenAI tools place on users, requiring heightened self-awareness, confidence adjustment, and flexible adaptation of problem-solving strategies~\cite{tankelevitch2024metacognitive}.

The idea that GenAI's ability to enhance productivity comes at the cost of diminished learning was further elucidated by researchers who implemented a GenAI assistant extension derived from StarCoder~\cite{li2023starcoder}. Their findings revealed that while GenAI-facilitated code completion substantially increased student productivity and operational efficiency, excessive dependence on these systems diminished creative problem-solving capacities and led to superficial comprehension of programming principles~\cite{takerngsaksiri2024students}. 
Despite these concerns, the prevailing consensus among both students and instructors is that these AI tools demonstrate substantial utility in addressing programming challenges within educational contexts~\cite{liffiton2023codehelp}, suggesting the need to balance productivity in the short term with learning in the long term. 
Some institutions have responded by fundamentally redesigning introductory courses around LLM integration, shifting emphasis from syntax mastery to skills like code explanation, testing, and problem decomposition~\cite{vadaparty2024cs1}.

% \vspace{-2mm}
\subsection{Student Interaction Patterns with GenAI Coding Tools}
Particularly relevant to our investigation is the 2024 study by Prather et al.~\cite{prather2024widening}, who conducted 21 laboratory sessions examining how novice programmers incorporate GenAI tools into their workflow when completing greenfield programming assignments. Their findings revealed a nuanced impact: while most students (20 out of 21) successfully completed their assigned problems, the benefits of GenAI assistance were unevenly distributed. Students who already possessed strong metacognitive skills achieved enhanced performance, while those with metacognitive difficulties were negatively impacted by GenAI use.

A separate investigation by Prather et al. ~\cite{prather2023s} further highlights the challenges faced by novice programmers when using Copilot. They discovered two novel patterns: novice programmers often (a) type code that matches Copilot's suggestions, but end up not accepting those suggestions (\textit{shepherding}); and (b) accept incorrect Copilot-generated code, leading to debugging rabbit holes that stray further from the correct solutions (\textit{straying}). Despite these challenges, participants believed that Copilot would always generate correct code. In a similar vein, Shah et al.~\cite{shah2025students} observed that students frequently employed "one-shot prompting," requesting Copilot to implement entire solutions at once, then dedicated subsequent interactions to debugging or regenerating incorrect responses. Notably, students reported greater trust in Copilot's code comprehension capabilities compared to its code generation features, attributing this difference to the presence of "trust affordances" in the Copilot chat interface that were absent in the code autocomplete functionality.
However, as observed in a study by Lauren E. Margulieux et al., Individual differences significantly influence how students interact with GenAI tools, with research showing that students with higher self-efficacy, lower fear of failure, and stronger prior academic performance tend to use AI less frequently or later in the problem-solving process~\cite{margulieux2024self}. These findings suggest that student characteristics may be more predictive of GenAI usage patterns than the tools' inherent capabilities.

Our study builds on this emerging body of research in two important ways: (a) by investigating the role of GenAI in brownfield programming tasks instead of greenfield programming tasks; and (b) by performing a detailed, second-by-second behavioral analysis of students' programming processes with and without the use of GenAI.

% \vspace{-2mm}
\section{Method}
\label{sections-s3-methods}
\subsection{Design}
We conducted a within-subjects, mixed-methods experimental study with two conditions corresponding to two different levels of the GenAI independent variable: \textsc{No Copilot} (control) and \textsc{Copilot} (experimental). In the \textsc{No Copilot} condition, participants worked on brownfield front-end web development tasks without the use of Copilot. Participants then switched to the \textsc{Copilot} condition, in which they worked on a set of highly similar front-end web development tasks with the use of Copilot. Because we were interested in first observing participants program without the use of Copilot, we did not counterbalance treatment order; all participants completed the \textsc{No Copilot} treatment first. However, we did counterbalance task order to mitigate task order effects.

Our study measured the impact of the independent variable on participants' task performance using two dependent variables: (a) time to complete tasks, and (b) number of tests passed. To characterize differences in participants' programming processes caused by manipulation of the independent variable, we performed a detailed categorization of participants' second-by-second programming behaviors. Finally, we elicited participants' perceptions of their programming processes through exit interviews.
\vspace{-2.5mm}
\subsection{Pilot Testing}
Prior to conducting the main study, we ran a pilot study with 13 participants to refine our experimental materials, tasks, and procedures. The pilot participants were recruited from the same population as our main study participants but were not included in the final sample. Pilot testing led to substantial revisions across multiple study components. We completely redesigned our programming tasks after finding that our initial search bar and sorting functionality tasks did not provide adequate similarity in complexity and scope. The replacement tasks—"Add Distance" and "Add Picture" features—offered better task equivalence while maintaining authentic brownfield programming challenges. We also enhanced our experimental rigor by developing robust and comprehensive automated test suites and refining our experimenter protocol. Post-interview questions were streamlined based on pilot feedback, and study timing was optimized to balance data collection thoroughness with participant engagement.

\vspace{-2.5mm}
\subsection{Participants}
Participants for this study were drawn from the undergraduate computer science program of Oregon State University, a large research university with over 4,000 students enrolled its undergraduate computer science programs. To recruit participants, we asked instructors of upper-division courses to post a study ad to their course webspaces. To incentivize participation, we offered participants a \$50 prepaid credit card for their participation. 

In this study, we were interested in exploring the impact of GenAI on undergraduate students with prior web development experience but little to no prior experience with GenAI programming assistants. To that end, we screened study applicants to ensure they (a) had little to no prior experience with using a GenAI programming assistant; (b) had completed at least one undergraduate course in front-end web development with an A or A-; (c) had completed at least one front-end web development project (at least three months in duration) outside the course; and (d) were in the third or fourth year of an undergraduate computing degree program.

We recruited 10 participants (9 male, 1 female, mean age 26.7): five traditional computer science majors and five post-baccalaureate computer science majors---three in their third year and seven in their fourth year. All participants demonstrated strong academic performance in a prior web development course, with 80\% having earned an A grade, and many having completed additional relevant coursework. Their practical web development experience varied: eight participants had created personal web projects, three had completed internships, and one possessed industry experience. Most participants (7) reported regular engagement with core web technologies, using JavaScript, HTML, and CSS at least weekly. Critically for this study, eight participants had no prior experience with GitHub Copilot, while the other two had only minimal exposure (used once or a few times).

\subsection{Materials and Tasks}
\subsubsection{Legacy Code Base}
Participants completed feature implementation tasks within a legacy code base consisting of 3,818 lines of code (49\% JavaScript, 40\% HTML, and 11\% CSS). Using plain vanilla JavaScript, the code base implemented an interactive front-end web application supporting a limited range of functionalities, including logging in, viewing and updating user settings, and creating, updating, and deleting user data. All user data was stored using \texttt{localStorage}, thus avoiding the need for the application to interact with a back-end database.

\subsubsection{Feature Implementations}
Participants in this study worked on two equivalent feature implementations within the context of the legacy code base---one in the \textsc{No Copilot} condition, and one in the \textsc{Copilot} condition. The feature implementations were carefully designed to possess highly similar properties, with comparable cognitive demands, technical complexity, and implementation scope. 

As shown in Table \ref{tab:feature-comparison}, both features required participants to implement new functionality across three sequential tasks with parallel structure: UI component construction (Task 1), interactive behavior implementation (Task 2), and data persistence operations (Task 3). The "Add Distance" feature required implementing the ability to enter a distance value into a field, with a toggle switch for unit conversion, while the "Add Picture" feature involved creating a profile picture selector with default and custom image options.

\begin{table*}[!t]
\caption{Comparison of 'Add Distance' and 'Add Picture' Features: Task Requirements, Implementation Details, and Testing Approaches}
\label{tab:feature-comparison}
\begin{tabular}{p{3cm}p{5.5cm}p{5.5cm}}
\toprule
\textbf{Feature} & \textbf{Add Distance} & \textbf{Add Picture} \\
\midrule
\textbf{Purpose} & 
Capture and validate user input for distance value and save it with user data. & 
Allow the user to select and save a profile picture (default or custom). \\
\addlinespace
\textbf{Tasks} & 
\begin{tabular}[t]{@{}l@{}}
1. Add Field, Radio Buttons\\
   and Label to Form\\
2. Convert Distances between\\
   Miles \& Kilometers\\
3. Save ``Distance" Field\\
   with other user data
\end{tabular} & 
\begin{tabular}[t]{@{}l@{}}
1. Add Profile Picture\\
   Selector to Form\\
2. Update Profile Picture\\
   based on Checked Status\\
3. Save Profile Picture\\
   in Local Storage
\end{tabular} \\
\addlinespace
\textbf{Input Components} & 
Distance input field, Radio buttons for unit selection. & 
Checkbox for custom picture selection, file picker and image preview \\
\addlinespace
\textbf{Default State} & 
Distance field empty, ``Miles" radio button selected. & 
File picker disabled, default image preview shown. \\
\addlinespace
\textbf{Interaction Flow} & 
Select unit with radio buttons, input distance, save distance with other data. & 
Select custom image, enable file picker, save chosen image. \\
\addlinespace
\textbf{Storage} & 
Distance saved in feet in Local Storage. & 
Profile picture saved in base64 in Local Storage. \\
\addlinespace
\textbf{Testing Approach} & 
Run a test suite to verify field functionality, unit conversions, and data storage. & 
Run a test suite to verify file picker behavior, image display, and data storage. \\
\bottomrule
\end{tabular}
\end{table*}

Quantitative analysis of the core properties of correct implementations of each feature confirmed their similar nature. As shown in Table ~\ref{tab:complexity-comparison}, both solutions required highly similar numbers of lines of code, program statements, variables, control structures, and operations. \footnote{Complete task specifications and problem statements are available at \url{https://ghcopilot-icer.github.io/}}

\begin{table}[!ht]
\caption{Implementation Complexity Analysis of Feature Tasks}
\label{tab:complexity-comparison}
\begin{tabular}{lcc}
\toprule
\textbf{Complexity Metric} & \textbf{Add Distance} & \textbf{Add Picture} \\
\midrule
Lines of code & 80 & 71 \\
Program statements & 29 & 28 \\
Variables & 4 & 4 \\
Control structures & 3 & 3 \\
Operations & 23 & 21 \\
\bottomrule
\end{tabular}
\end{table}

\subsubsection{Test Suites}
We developed automated test suites to verify the correct behavior of solutions to the three tasks within each feature. For both features, Task 1 contained four test cases, Task 2 contained four test cases, and Task 3 contained five test cases. The test cases verified minimum functional UI rendering, behavioral correctness, and data persistence for each feature, providing consistent and objective verification criteria.

\subsubsection{Participant Instructions}
At the start of the study, participants were asked to read aloud a set of written instructions that (a) indicated that two hours were allocated for the study; (b) described the purpose of the study, (c) reminded participants that GitHub Copilot, and not them, was being tested, (d) provided an overview of what participants would be doing in the session; and (e) instructed participants to think aloud during the study.

\subsubsection{Tutorial}
We designed a tutorial to introduce participants to (a) the web application they would be enhancing with new features, (b) the application's code base, and (c) GitHub Copilot. The tutorial began by having participants launch the application in a web browser. The tutorial instructed participants to log in, and then directed participants to explore the app pages (user settings and user data) that they would be modifying within the study. Finally, the tutorial walked participants through a warm-up task in which they added an "About" dialog box to the application. Participants were instructed to use Copilot for the first two steps of the warm-up task, but then to switch Copilot off for the remaining steps. In the Copilot section of the tutorial, participants received explicit instructions on how to use the two core features of Copilot: (a) the \textit{code completion feature}, which generates code suggestions (in \textit{\textcolor{gray}{gray italics}}) at the current editing insertion point; and (b) the \textit{chat facility}---a side panel for conversing with the GenAI chatbot. We demonstrated both of these features by having participants type in a short code snippet and select the Copilot autocompletion, and by having participants enter a sample prompt into the chat facility. However, we provided no further advice on their use. Additionally, Copilot offers a third method of directly inserting code from Copilot chat that was introduced in November 2024, but this feature was not included in our tutorial as it was introduced after the study design was finalized.

\subsubsection{Task Description Documents}
We developed documents for the \textsc{No Copilot} and \textsc{Copilot} versions of the two highly similar feature implementations. Both the \textsc{No Copilot} and \textsc{Copilot} documents (a) presented screenshots of the feature to be implemented within the context of the existing application page; (b) instructed participants that they were free to use the web to search for help or solutions; (c) instructed participants that they could use Copilot to complete the tasks (in the \textsc{No Copilot} treatment, the Copilot extension was disabled); (d) identified the project files that participants would need to edit to complete the implementation; (e) told participants that the application uses \texttt{localStorage} instead of a back-end database for storage; (f) identified specific solution requirements for each task within the feature implementation (e.g., ``A `distance field' must appear centered in the form below the `score' field''); (g) instructed participants to execute the test suite at any time to test their code; and (h) indicated that they were done with a task when all tests in the test suite pass.

\subsubsection{Experimental Computing Environment}
Participants completed study sessions by logging into an Amazon Web Services Workspace~\cite{amazon_amazon_nodate}---a Linux-based virtual machine configured with a minimal set of software, accounts, and files necessary for study participation:
\begin{itemize}
    \item Visual Studio Code IDE~\cite{vscode} with the Copilot Extension~\cite{copilot}, which participants used to work on all programming tasks
    \item Google Chrome web browser, which participants used to view the study documents and perform web searches during programming tasks
    \item Node.js, the build environment for the web application that participants worked on in the study
    \item Git, for committing code changes, with the user logged in under an anonymous Git account to protect their identity
    \item a fresh version of the web application source code, under Git source control, with a separate remote repository for each participant
    \item Participant instructions, tasks, and tutorial documents
\end{itemize}

\subsection{Procedure}
Because we did not want Copilot to influence participants' initial task performance, we had all participants complete the \textsc{No Copilot} treatment first, followed by the \textsc{Copilot} treatment. However, we fully counterbalanced the order of the two feature implementations to control for task order and task-treatment effects. 

Prior to the study sessions, students interested in participating in the study completed a screening questionnaire in which they were asked about their prior web development experience and prior use of GenAI coding assistants. Based on their screening questionnaire responses, eligible participants were emailed a link to book a study session slot.

Study sessions lasted 120 minutes and were conducted via Zoom, with recording turned on, but with participants' screen names anonymized (e.g., ``P01'') and participants' cameras turned off to protect their identity. At the start of each session, an experimenter greeted the participant and reconfirmed that they did not have experience with a GenAI coding assistant. The participant was then given a link to our informed consent form. If they consented to participate, they were redirected to a brief online demographic questionnaire. 

The participant was then directed to download and install on their local computer the AWS Workspace client, which they needed to access the experimental computing environment. Once logged into that environment, the participant shared their screen and performed a speed test to verify they had a sufficiently fast internet connection to complete the study.

After reading the task instructions aloud, the participant was given 15 minutes to complete the tutorial. The participant then worked on the two feature implementations. For each task within a feature, the participant
\begin{itemize}
    \item read the task instructions aloud
    \item implemented the code to address the task
    \item ran (at any time) automated tests to verify their implementation
    \item committed their code when the task implementation was complete
\end{itemize}

Participants were given up to 30 minutes to implement each feature, inclusive of testing activities. However, we excluded the time spent on the initial read-through of task instructions from the 30-minute limit. Upon successful completion of individual tasks within a feature (as confirmed by passing all test cases), we paused the timer while participants read instructions for the next task. Timing resumed only after participants indicated their readiness to begin the next task. This enabled us to measure actual implementation time consistently for all participants.

\subsection{Data Collection and Analysis}

We employed a mixed-methods approach to collect and analyze participants' coding outcomes, processes, and perceptions. All participant sessions were video recorded and subsequently analyzed to extract both quantitative and qualitative data, including (a) time to complete each task, (b) number of test cases passed, (c) categorical coding of participants' programming behaviors, and (e) transcriptions of the exit interviews. In addition, each participant committed their code solutions to a GitHub repository. We used these repositories to cross-check the video data on test cases passed by executing participants' code solutions against the automated test suites.

%Qualitative data was gathered concurrently through three channels. Participants engaged in think-aloud protocols during task completion, completed background questionnaires, and participated in structured exit interviews. These exit interviews, conducted at the conclusion of each session, elicited participants' overall experiences and perspectives, enriching our understanding of their coding activities.

\subsubsection{Programming Process Coding Scheme}
To systematically analyze participants' programming behaviors with and without GitHub Copilot, we developed a fine-grained activity coding scheme that captured both high-level activities and specific interactions with Copilot. This scheme enabled quantitative analysis of how participants allocated their time across different programming activities and how these patterns changed when using GenAI assistance.

Our coding scheme emerged through a bottom-up, inductive approach based on direct observation of programmer behavior. The research team analyzed video recordings of programming sessions, identifying distinct activities and interaction patterns as they naturally occurred. Through collaborative discussion, we established an initial set of activity categories encompassing the full range of observed behaviors, then iteratively refined category definitions and boundaries through additional video analysis.

The scheme includes both primary and secondary categories (see Table~\ref{tab:annotation-schema}). Primary categories represent distinct high-level activities such as writing code, viewing code, and testing, while secondary categories capture more specific behaviors within these activities, such as accepting Copilot suggestions or manually editing code.

\begin{table*}
\caption{Coding Scheme for Programming Activities}
\label{tab:annotation-schema}
\rowcolors{2}{white}{gray!15} % Start alternating colors from row 2 with white and light gray
\begin{tabular}{p{3cm}p{4cm}p{7cm}}
\toprule
\textbf{Primary Category} & \textbf{Secondary Category} & \textbf{Description} \\
\midrule
\rowcolor{white} % Explicitly set color for the multirow row
\multirow{1}{*}{View Task} & -- & Searching or viewing task materials and instructions \\

\rowcolor{gray!15} % Explicitly set color for the multirow row
\multirow{1}{*}{View Code} & -- & Searching or viewing code in editor or file explorer \\

\rowcolor{white} % Explicitly set color for the multirow row
\multirow{1}{*}{View Web} & -- & Searching or viewing web materials for reference \\

\rowcolor{gray!15} % Explicitly set color for the multirow row
\multirow{1}{*}{View App} & -- & Viewing the web application in browser \\

\rowcolor{white} % Explicitly set color for the multirow row
\multirow{1}{*}{View Dev Tools} & -- & Viewing browser developer tools \\

\rowcolor{gray!15} % Explicitly set color for the multirow row
\multirow{1}{*}{View Response} & -- & Viewing Copilot chat responses or autocomplete suggestions \\

\rowcolor{white} % Explicitly set color for the multirow row
\multirow{-1}[0]{*}{Write Code} & Paste Copilot & Copy/paste code from Copilot chat \\
\rowcolor{white} & Accept Copilot Suggestion & Accept Copilot autocomplete suggestion (via tab) \\
\rowcolor{white} & Insert using Copilot& Insert code using inline Copilot chat interface (Copilot Edits) \\
\rowcolor{white} & Paste Web & Copy/paste code from web sources \\
\rowcolor{white} & Enter/Edit & Manually enter or edit code \\
\rowcolor{white} & Paste Self & Copy/paste from current code project \\
\rowcolor{white} & Modify Copilot Suggestion & Modify or delete recently accepted Copilot suggestion \\

\rowcolor{gray!15} % Explicitly set color for the multirow row
\multirow{-1}[0]{*}{Test CLI} & Run Test & Running tests from command line \\
\rowcolor{gray!15} & View Test & Viewing test results \\

\rowcolor{white} % Explicitly set color for the multirow row
\multirow{-1}[0]{*}{Interacting } & Resolve Question & Resolving questions with experimenter \\
\rowcolor{white} & Prompt Participant & Experimenter prompting participant \\

\rowcolor{gray!15} % Explicitly set color for the multirow row
\multirow{-1}[0]{*}{Write Prompt } & Paste Task & Copy/paste from task materials into prompt \\
\rowcolor{gray!15} & Paste Previous Copilot Output & Copy/paste from previous Copilot response \\
\rowcolor{gray!15} & Paste Error & Copy/paste from error messages \\
\rowcolor{gray!15} & Enter Prompt & Manually enter/edit prompt \\
\rowcolor{gray!15} & Enter Prompt as Comment & Enter prompt as code comment \\

\rowcolor{white} % Explicitly set color for the multirow row
\multirow{1}{*}{Idle} & -- & No active interaction occurring \\
\bottomrule
\end{tabular}
\end{table*}

For each category, we developed precise start and stop criteria to ensure consistent application of the coding scheme. For example, Write Code (WC) starts when participants begin typing in the code editor and stops when they shift focus to viewing code, running tests, or other activities. Secondary categories provide additional granularity, particularly for code writing, where we distinguish between manually entered code, accepted Copilot suggestions, pasting Copilot responses, and other interactions.

To verify the reliability of our coding scheme, we conducted a rigorous analyst training and evaluation process. Through this process, we iteratively refined category definitions and start/stop criteria until reliable agreement was achieved. Three analysts independently coded 15\% of the total video data (90 minutes out of 10 hours) following the coding guidelines. We then calculated inter-rater reliability using Krippendorf's Alpha\cite{krippendorff2011computing} with a time-windowed approach, analyzing agreement in one-second intervals. This yielded Krippendorf's alpha values of 0.91 for primary categories and 0.85 for secondary categories. Having established sufficiently high inter-rater reliability, the three analysts divided up the coding of the remaining video data.

\section{Results}
\label{sections-s4-results}
%Our analysis revealed significant differences in programming behaviors, performance, and workflows We present our findings organized by research question, focusing on performance impacts, interaction patterns, and student perceptions.

\subsection{Impact on Task Performance (RQ1)}
\subsubsection{Task Efficiency}

\begin{figure}[!ht]
\centering
\begin{subfigure}[b]{0.45\textwidth}
\includegraphics[width=\textwidth]{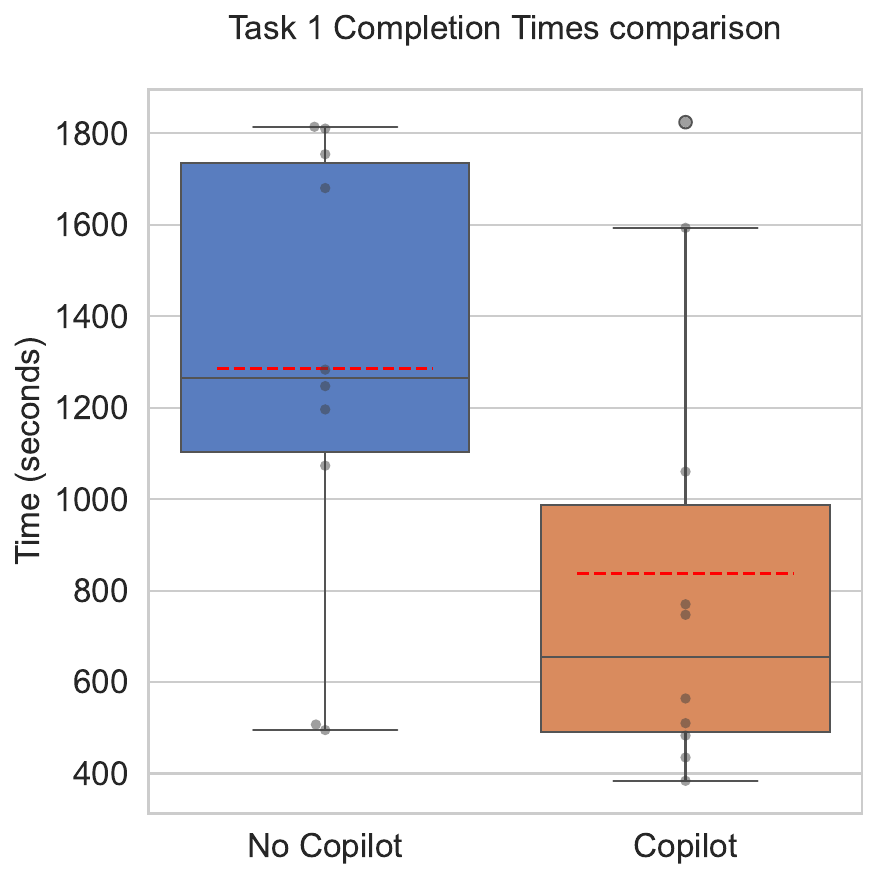}
\caption{Box plots of time to complete Task 1 by condition  (p = 0.037)}
\end{subfigure}
\hfill
\begin{subfigure}[b]{0.45\textwidth}
\includegraphics[width=\textwidth]{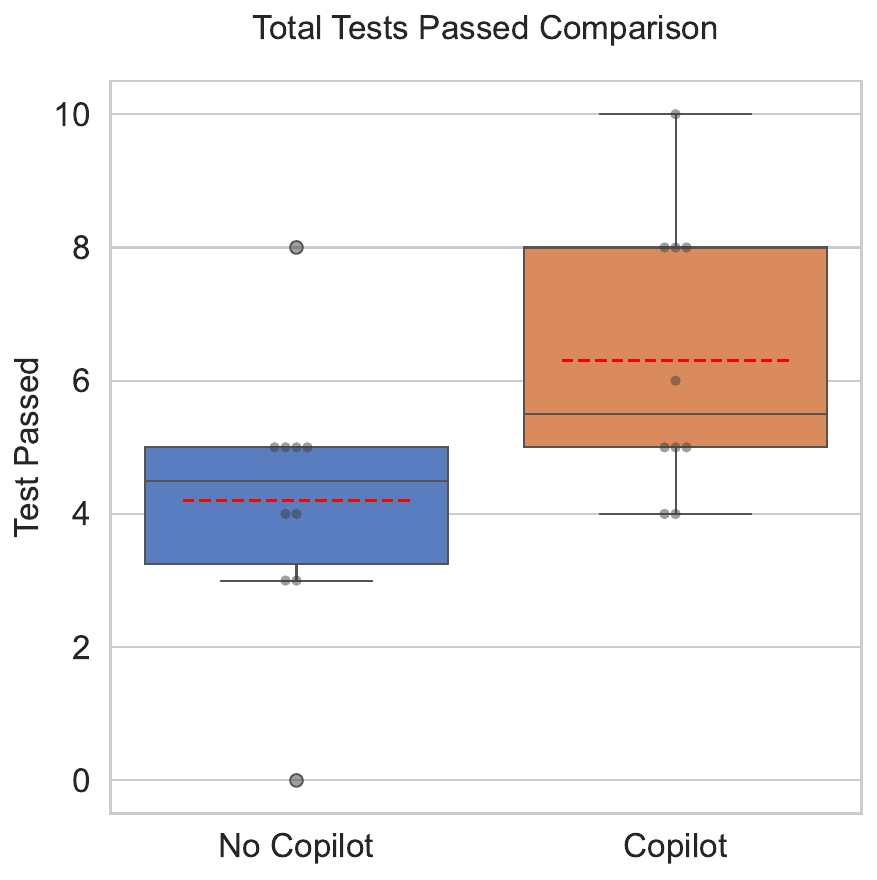}
\caption{Box plots of number of tests passed by condition (p = 0.012)}
\end{subfigure}
\caption{Box plots of participant task performance with and without Copilot (\textit{n}=10). Red dotted line shows the mean value.}
\label{fig:task-completion-tests}
\end{figure}

Figure \ref{fig:task-completion-tests}(a) presents a box plot of the time participants needed to complete Task 1 (the only task all participants completed) with and without Copilot. Participants completed this task in an average of 34.9\% less time with Copilot (\textit{M} = 837.0 seconds, \textit{SD} = 496.1 seconds) than without it (\textit{M} = 1285.9 seconds, \textit{SD} = 503.5 seconds). This difference is statistically significant (\textit{W} = 7.0, \textit{p} = 0.04), with a large effect size (\textit{r} = 0.66).

This quantitative finding was strongly supported by participants' subjective experiences. As S05 explained, ``I'm probably like 4 to 5 times more productive, or maybe even more, with Copilot than I am... just using Google searching.'' Similarly, S11 contrasted the approaches: ``Without [Copilot]... it was kind of a lot slower. I had to look up some things and double-check my answer. It took me 20 minutes or so... with GitHub Copilot, I was able to finish a more complex task in just a few seconds.''

\subsubsection{Solution Completeness and Correctness}
Figure \ref{fig:task-completion-tests}(b) presents a box plot of the number of tests that participants passed with and without Copilot. When working with Copilot, participants passed an average of 6.30 automated tests (\textit{SD} = 1.57), compared to 4.20 tests (\textit{SD} = 1.87) without Copilot. This difference is statistically significant (\textit{W} = 1.50, \textit{p} = 0.01), with a large effect size (\textit{r} = 0.80). Given that passing a test indicated that a single task requirement had been correctly addressed, this finding has two implications: (a) Copilot enabled participants to make significantly more progress toward implementing features; and (b) Copilot led to significantly more correct (partial) solution implementations.

Participants' reflections provided insight into this improvement. S01 noted that with Copilot, ``I felt like I had a lot of time to do those fine-tuning details,'' whereas without it, they ``had to spend quite a bit of time to set up all the HTML and JavaScript.'' This shift in focus from implementation to refinement appears to have helped students make more progress toward correctly implementing new features.

\subsection{Programming Processes (RQ2)}
\subsubsection{Programming Activity Distribution}

\begin{figure*}[!ht]
    \centering
    \includegraphics[width=1\linewidth]{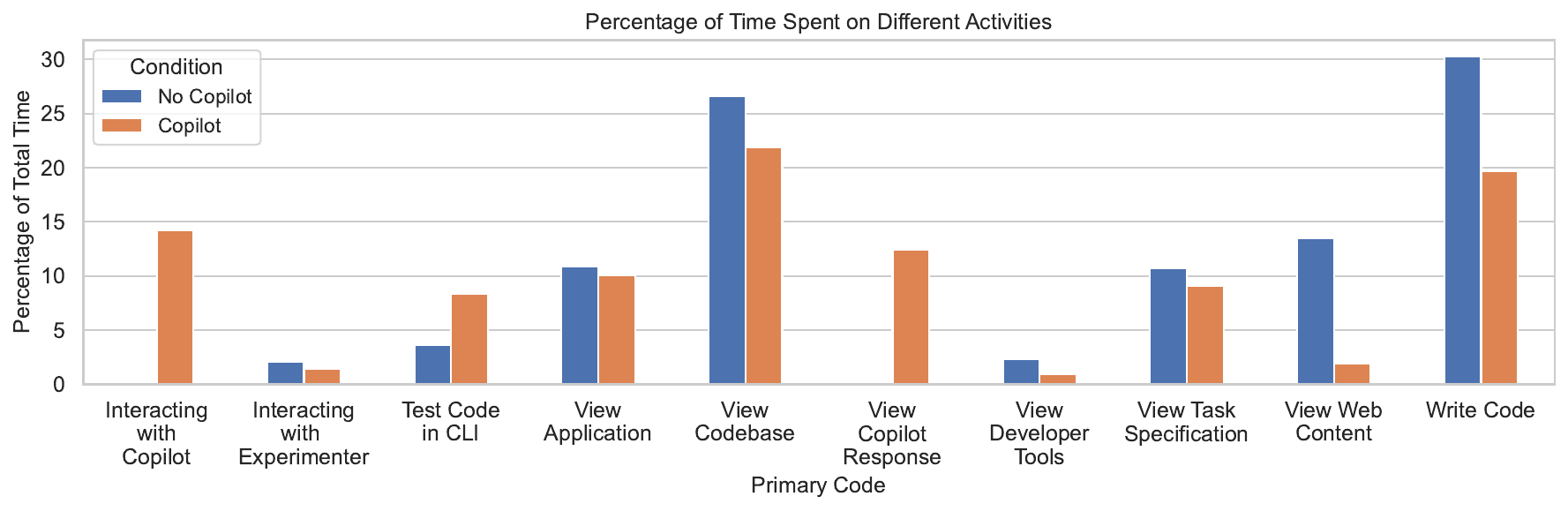}
    \caption{Interaction pattern in engaging different activities while solving the problem}
    \Description{Comparison of Programming Activities with and without Copilot}
    \label{fig:code-writing-shift}
\end{figure*}

Figure \ref{fig:code-writing-shift} presents a bar graph comparing the percentages of task time participants spent performing various programming activities with and without Copilot. To test for differences in the overall distribution of percentages of time spent in each category, we performed a permutation-based chi-square test~\cite{pesarin2010permutation}, which preserves the within-subjects nature of the comparison. The test revealed a significant difference in the categorical distributions of programming activities in the \textsc{Copilot} and \textsc{no-Copilot} conditions ($\chi^2 = \text{42.01}, df = \text{9}, p = \text{0.004}$, based on 10,000 permutations), with a small effect size (Cramer's V = 0.18). 

\begin{table*}[!htbp]
\caption{Comparison of Activity Distribution With and Without Copilot. WP (Write Prompt), VW (View Web), VR (View Response), and WC (Write Code) remained significant after BY adjustment.}
\label{tab:activity-comparison}
\begin{tabular}{lccccc}
\toprule
\textbf{Category} & \textbf{Without Copilot (\%)} & \textbf{Copilot (\%)} & \textbf{W-Stat} & \textbf{\textit{p}-value} & \textbf{Effect Size (\textit{r})} \\
\midrule
Write Prompt (WP) & 0.00 & 14.27 & 0.0 & 0.004*  & 0.96 \\
View Response (VR) & 0.00 & 12.45 & 0.0 & 0.004*  & 0.96 \\
View Web (VW) & 13.51 & 1.91 & 0.0 & 0.004* & 0.96 \\
Write Code (WC) & 30.45 & 19.82 & 1.0 & 0.008*  & 0.89 \\
View Code (VC) & 26.56 & 21.95 & 14.0 & 0.359  & 0.31 \\
Test CLI (TC) & 3.75 & 8.31 & 10.0 & 0.164  & 0.46 \\
View Dev Tools (VD) & 6.83 & 2.64 & 0.0 & 0.250  & 0.66 \\
View Task (VT) & 10.61 & 9.03 & 13.0 & 0.301  & 0.34 \\
View App (VA) & 10.78 & 9.99 & 15.0 & 0.426  & 0.27 \\
Interacting (IN) & 2.33 & 1.56 & 12.0 & 0.461  & 0.26 \\
\bottomrule
\end{tabular}
\begin{center}
\begin{tablenotes}
\small
\item Values represent percentage of time spent on each activity. W-Stat = Wilcoxon statistic.
\item * Significant at $\alpha$ = 0.05 after Benjamini-Yekutieli correction.
\end{tablenotes}
\end{center}
\end{table*}

To probe for where the pair-wise categorical differences lie, we applied post-hoc Wilcoxon signed-rank tests. Statistical significance was determined using the Benjamini-Yekutieli procedure~\cite{benjamini2001control} to control the false discovery rate at $\alpha = 0.05$ across multiple comparisons. Table~\ref{tab:activity-comparison} presents a summary of individual categorical differences, the results of the statistical tests, and the effect sizes. As the table indicates, there were four statistically significant behavior shifts, all of which had large effect sizes:

\begin{itemize}
    \item With Copilot, participants spent significantly less time writing code.
    \item With Copilot, participants spent significantly less time viewing web content.
    \item With Copilot, participants spent significantly more time writing prompts and viewing Copilot responses. These differences were expected, given that Copilot was not available in the \textsc{No Copilot} condition.
\end{itemize}

% To probe for where the pair-wise categorical differences lie, we applied post-hoc Wilcoxon signed-rank tests. Statistical significance was determined using the the Benjamini-Yekutieli (BY) procedure~\cite{benjamini2001control} to control the false discovery rate at $\alpha = 0.05$ across multiple comparisons. Table~\ref{tab:activity-comparison} presents a summary of individual categorical differences, along with the results of the statistical tests. After applying the BY correction, only one activity showed a statistically significant difference:

% \begin{itemize}
%     \item With Copilot, participants spent significantly less time writing code.
% \end{itemize}
% While not reaching significance after the strict BY correction, several other notable shifts with large effect sizes were observed:
% \begin{itemize}
%     \item With Copilot, participants spent significantly less time viewing web content.
%     \item With Copilot, participants spent significantly more time writing prompts and viewing Copilot responses. These differences were expected, given that Copilot was not available in the No Copilot condition.
% \end{itemize}

These results, which suggest a clear shift from implementation to verification activities, align with participants' self-reported experiences. S10 described traditional programming as ``a lot of jumping back and forth between writing the code and searching,'' while observing that with Copilot, ``I didn't have to make back and forth jumps between Google.'' Similarly, S05 emphasized that Copilot eliminated the need to ``go to Stack Overflow or MDN... and not even have to read the documentation or find what you're looking for.'' Similarly, S03 observed that, without Copilot, ``it is a little harder... I have to figure out which sources to use when I googled the question. It is hard to find the correct answer.'' These comments suggest that Copilot serves as an integrated knowledge source for students, reducing context switching and the need for external information seeking.

% \subsubsection{Web Search Behavior}

% \begin{figure}[!ht]
%     \centering
    
%     \begin{subfigure}[b]{0.45\textwidth}
%         \includegraphics[width=\textwidth]{figures/web_search_count.pdf}
%     \end{subfigure}
%     \begin{subfigure}[b]{0.45\textwidth}
%         \includegraphics[width=\textwidth]{figures/web_search_total_duration.pdf}
%     \end{subfigure}
%     \caption{Web search behavior comparison showing (a) number of web searches and (b) total time spent on web searches with and without Copilot. The 69\% reduction in search time was statistically significant (p=0.034).}
%     \label{fig:web-search-behavior}
%     \Description{Web search behavior comparison showing (a) number of web searches and (b) total time spent on web searches with and without Copilot. The 69\% reduction in search time was statistically significant (p=0.034).}
% \end{figure}

% The data reveals important differences in web search behavior when using Copilot ~\ref{fig:web-search-behavior}. Most notably, developers spent 69\% less time conducting web searches(without copilot 228.11 seconds on average with 130.28 standard deviation while with copilot 70.75 seconds on average with 75.92 standard deviation) (p=0.034). The number of web searches conducted also decreased substantially (3.5 searches on average with 2.52 standard deviation vs. 8.56 searches on average with 4.48 standard deviation) when using Copilot, a 59\% reduction that approached statistical significance (p=0.051).

\subsubsection{Code Writing Behavior}
\begin{figure*}[!ht]
    \centering
    \includegraphics[width=0.9\linewidth]{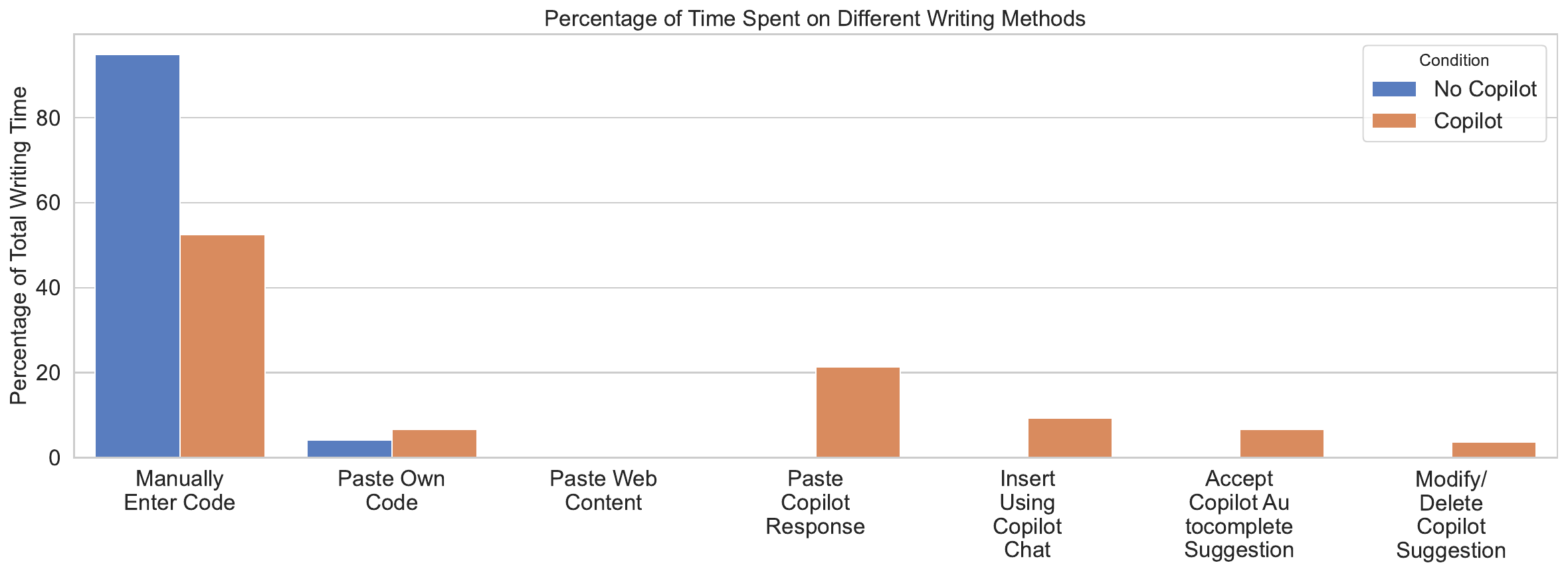}
    \caption{Breakdown of code writing activities, showing the  shift from almost exclusively manual code entry without Copilot to a  mix of coding methods with Copilot assistance.}
    \Description{Breakdown of code writing activities showing a  shift from almost exclusively manual code entry without Copilot to a mix of coding methods with Copilot assistance.}
    \label{fig:code-writing-behavior}
\end{figure*}

Recall that we defined a secondary coding scheme to capture participants' code writing behaviors at a finer level of granularity. Figure~\ref{fig:code-writing-behavior} compares participants code writing behaviors in the \textsc{No Copilot} and \textsc{Copilot} conditions based on the secondary categories. According to a permutation-based chi-square test, there was a statistically significant difference between the coding activities of the \textsc{No Copilot} and \textsc{Copilot} conditions ($\chi^2\ = 171.63, df = \text{7}, p = \text{0.003}$, based on 10,000 permutations), with a large effect size (Cramer's V = 0.70). Without Copilot, participants relied almost exclusively on manual code entry (95.6\% of coding time). With Copilot, manual coding dropped to 52.5\% of coding time and was augmented with Copilot assistance: pasting Copilot responses (21.3\%), inserting using Copilot (9.3\%), accepting suggestions (6.6\%), and modifying suggestions (3.7\%).

This shift in code-writing behavior was reflected in participants' descriptions of their process. S07 noted that ``it made a lot of tasks easier,'' while S08 appreciated that ``suggestions are what I was going to write out and saved time.'' However, this transformation came with challenges, as S04 observed: ``Since you don't write yourself, it feels very unorganized. You don't know where to put it.''

\subsubsection{Workflow Transitions}

\begin{figure}
    \centering
    \begin{subfigure}[b]{0.45\textwidth}
        \includegraphics[width=\textwidth]{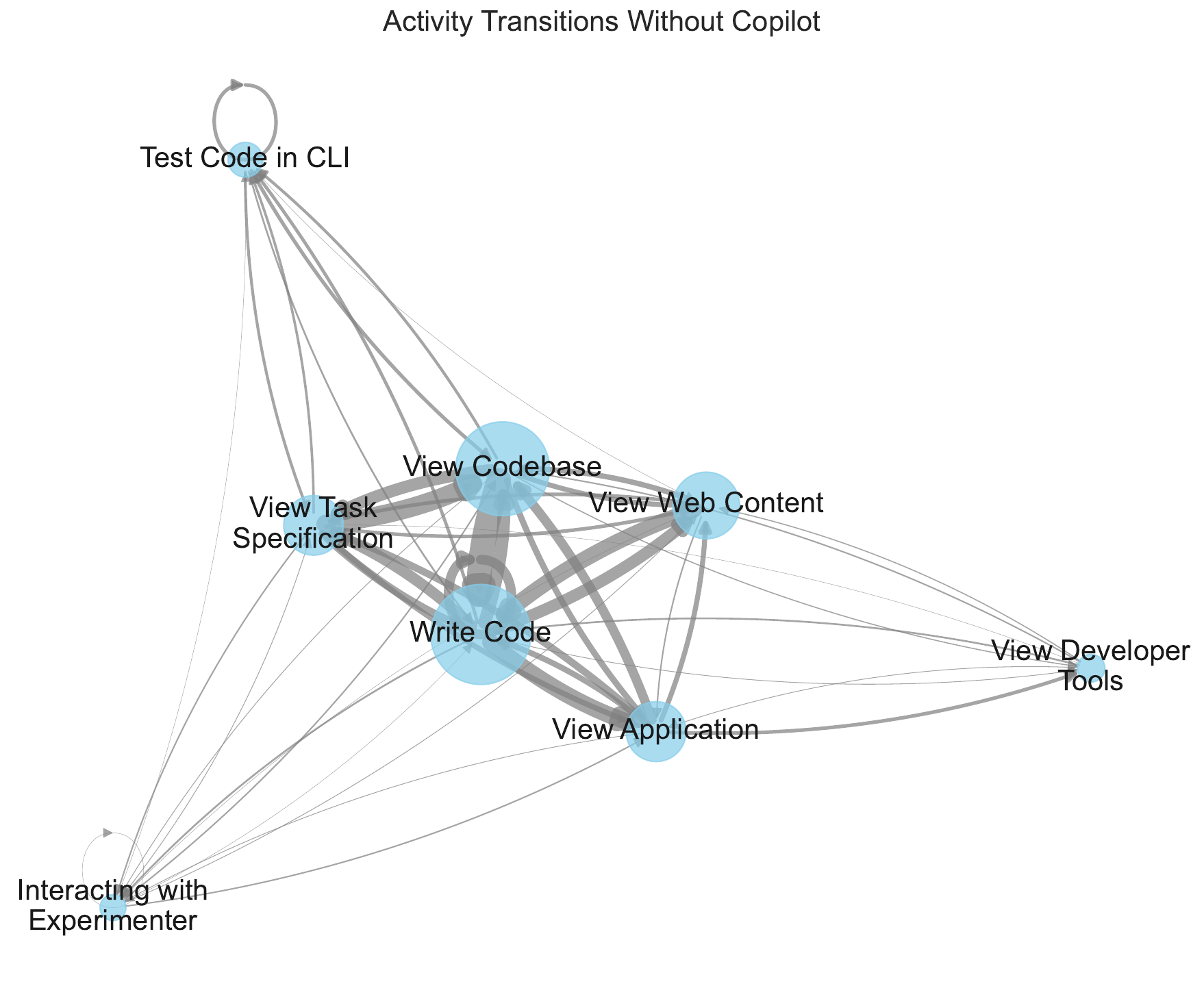}
    \end{subfigure}
    \begin{subfigure}[b]{0.45\textwidth}
        \includegraphics[width=\textwidth]{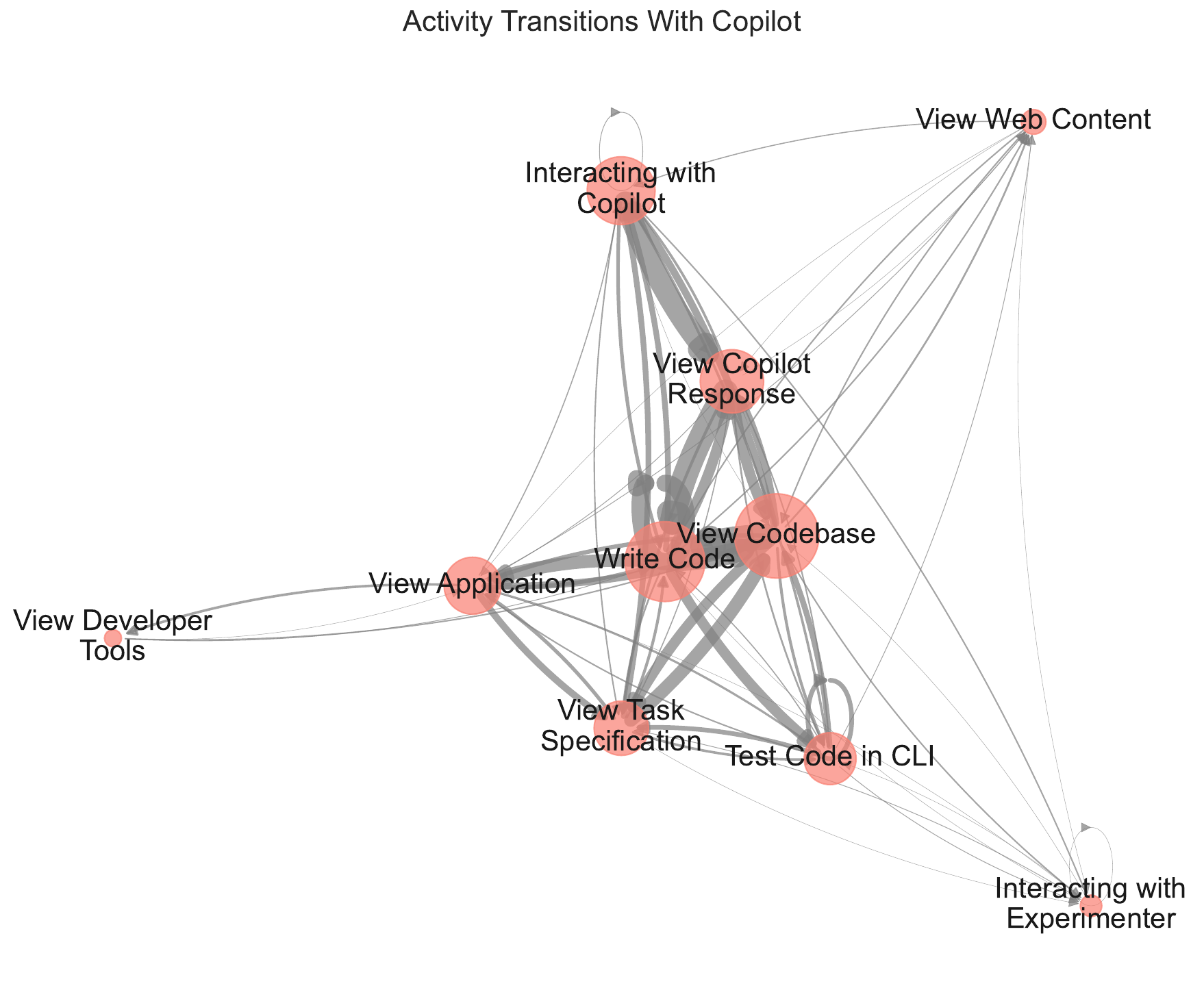}
    \end{subfigure}
    \caption{Programming workflow networks comparing activity transitions without Copilot (top) and with Copilot (bottom). Node size indicates time spent on each activity, while arrow thickness shows how frequently students transitioned between activities. The layout positions frequently connected activities closer together, revealing the emergence of a GenAI-mediated prompt→response→implement cycle with Copilot.}
    \Description{Workflow transition diagrams showing the emergence of a  GenAI-mediated coding cycle with Copilot (right) compared to a traditional development workflow without Copilot (left). Arrow thickness represents transition frequency, while node proximity indicates activity co-occurrence, with adjacent nodes representing frequently sequential activities and distant nodes representing less common transitions.}
    \label{fig:markov-transition-behavior}
\end{figure}

To understand differences in participants' workflow patterns, we performed a Markov transition analysis~\cite{visser2010stochastic} by calculating the probabilities of transitions between sequential activities. Figure ~\ref{fig:markov-transition-behavior} visualizes these workflow patterns, revealing stark differences. Most notably, a GenAI-mediated coding cycle emerged: \textit{prompt→~view response→~implement}, with transitions between these states occurring at high frequencies (e.g., \textit{prompt~→~read response}: 78.21\%). This cyclic pattern was absent in the No Copilot condition, where developers instead followed a traditional \textit{read~→~understand~→~implement} workflow.

A permutation test (10,000 iterations) comparing the Markov transition matrices between conditions revealed significant differences in workflow patterns (sum of squared differences = 1.86, p < 0.0001), confirming that Copilot fundamentally restructured participants' programming activities. The largest differences occurred in transitions from viewing development tools (Kullback-Leibler divergence =5.48) and from viewing web resources (Kullback-Leibler divergence=3.17). This suggests that Copilot reduced participants' reliance on external resources for information seeking. Additionally, the testing activity appears more integrated into the overall workflow with Copilot, showing increased connections to various activities including the GenAI-mediated cycle. This integration of testing is consistent with the finding that participants shifted toward more verification activities when using Copilot.

S09 described this workflow transformation as ``very streamlined as far as the small tasks I needed to accomplish,'' while S10 noted that when Copilot ``gets it right, I just hit tab, and it autocompletes it... it helps me save a lot of time in terms of typing and syntax.''

\subsection{Copilot Usage Strategies (RQ3)}
To explore whether participants' task success might be related to their Copilot usage strategies, we partitioned the participants into "lower" and "higher" performing groups based on the total number of test cases they passed in the two conditions. There was a clear gap between the fourth-highest ranked participant (13 test cases passed) and the fifth-highest ranked participant (9 test cases passed), providing a natural dividing line for splitting participants into "lower performers" (six participants) and "higher performers" (four participants).

\begin{figure*}
    \centering
    \includegraphics[width=0.75\linewidth]{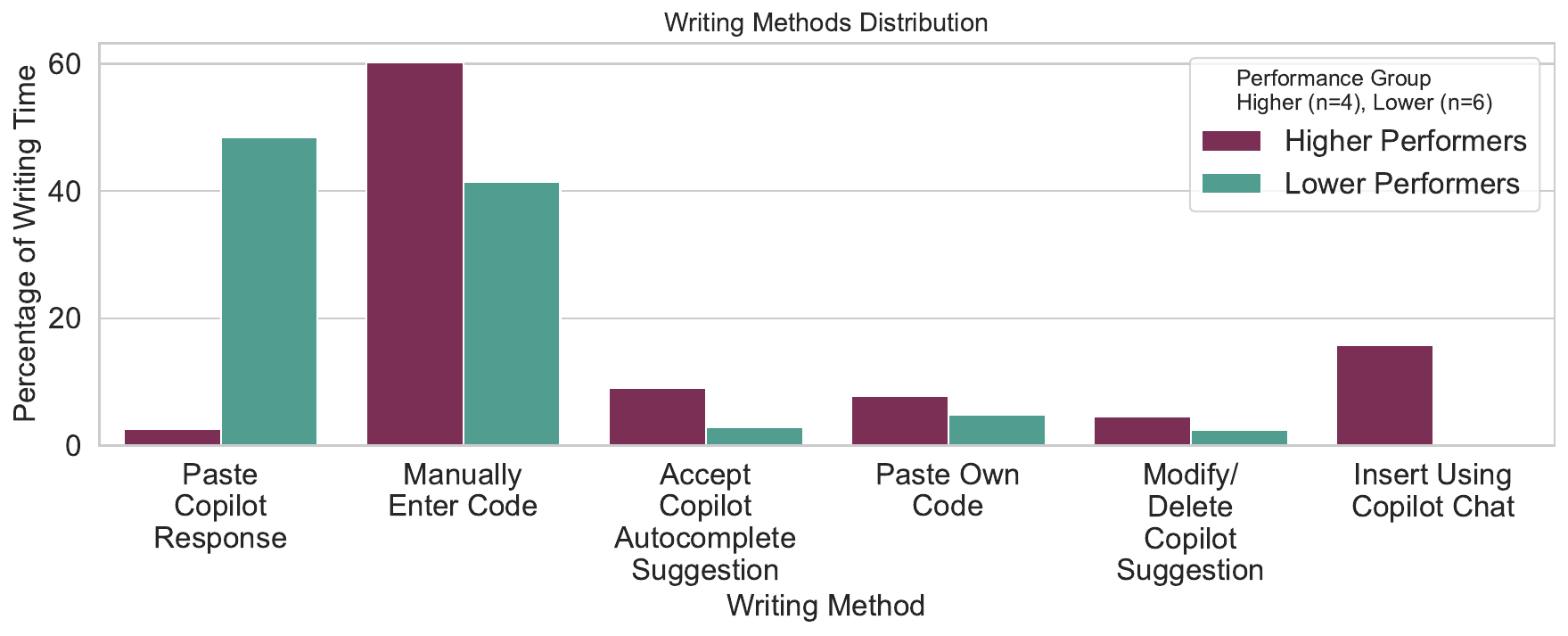}
    \caption{Comparison of the Copilot interaction strategies of higher and lower performers}
    \label{fig:writing-method-distribution}
\end{figure*}

Figure~~\ref{fig:writing-method-distribution} presents a comparison of the lower and higher performing groups relative to their code writing behaviors in the Copilot condition. A chi-square test of independence revealed significant differences in how high and low performers utilized Copilot ($\chi^2\ = 56.16, df = \text{7}, p < \text{0.0001}$). Analysis of standardized residuals identified that the only statistically significant difference was that low performers spent significantly more of their code writing time directly pasting code from the Copilot chat, compared to high performers (\textit{p} < 0.001). While not reaching statistical significance, additional trends suggest that high performers tended to spend higher percentages of their implementation time accepting Copilot suggestions and pasting their own code. These patterns suggest that higher-performing participants may engage more selectively and interactively with Copilot than lower-performing participants, who tended to use chunks of Copilot code wholesale. However, further research with larger samples would be needed to confirm these trends.

% A qualitative analysis revealed varying strategies for integrating Copilot into brownfield programming tasks. One notable approach was taken by Participant S11 (a high performer), who completed Task 2 with 100\% accuracy in just four minutes by providing the entire problem statement to Copilot Chat. Other participants employed more incremental strategies, combining manual implementation with integration of targeted Copilot suggestions.

%Though further research with larger samples is needed to confirm these results, they suggest a key difference in how the lower- and higher- performing participants appropriated Copilot: Whereas lower-performing participants were more likely to use chunks of Copilot code wholesale, higher-performing participants tended to incorporate Copilot-generated code at a more granular level, combining Copilot autocomplete suggestions with manual code editing. This suggests that the higher-performing participants were more selective about the Copilot-generated code they used---a possible sign of increased metacognitive engagement with the tool. 

% Interestingly, the analysis of Copilot interaction patterns revealed a critical distinction between high and low performers ~\ref{fig:writing-method-distribution}. Top-performing participants pasted Copilot code 83.9\% less frequently than bottom performers (p=0.045), instead showing a 400\% increase in accepting inline suggestions. This suggests that more selective, granular incorporation of AI-generated code may be more effective than wholesale adoption of larger code blocks.

Participants reported mixed experiences regarding how Copilot affected their brownfield coding activities. S03 (a lower performer) found that with Copilot, ``it is easier to find out where is supposed to be editing,'' while S07 (a lower performer) noted that "it's easier to familiarize yourself" with the code base using Copilot. However, others like S11 (a higher performer) observed that Copilot ``didn't help me understand necessarily'' because ``it just kind of gave me the answer.``
This tension between productivity and understanding was evident in several responses. S04 (a lower performer) reflected: ``I feel like it's much easier to understand code, or Copilot forces you to kind of read the code in a way,'' while also noting challenges with integration: ``It's not like you are thinking, it's not your own logic. So you don't know where it goes.''
S05 (a lower performer) expressed a similar concern: ''If you don't understand the code base... you're probably even a little bit better off without Copilot, because you're thinking more about what needs to go where.'' These participant perceptions suggest that while Copilot enhances the productivity of student programmers in brownfield coding tasks, productivity gains may come at the cost of a diminished understanding of the legacy code base and how GenAI-generated code fits into the code base.  %This highlights a complex relationship between GenAI assistance, productivity, and understanding in brownfield programming contexts—--a critical consideration for future computing education research.

\section{Discussion and Implications}
\label{sections-s5-discussion}
% Our study investigated the impact of GitHub Copilot on student programmers working with legacy codebases. The findings revealed significant changes across multiple dimensions of programming behavior and performance. We structure our discussion around our research questions, examining both the quantitative improvements and the qualitative insights that emerged from our mixed-methods analysis.

\subsection{RQ1: Copilot's Impact on Programming Performance}

Our results demonstrate that third and fourth-year undergraduate students complete brownfield programming tasks more efficiently, completely, and correctly with the help of GenAI. The 34.9\% reduction in task completion time and 26.2\% improvement in test pass rates align with previous studies of the impact of GenAI programming assistants on both professional programmers ~\cite{peng2023impact, ng2024harnessing} and undergraduate students~\cite{frank2024enhancing, prather2024widening} in greenfield programming scenarios. Our study extends these previous results by demonstrating that GenAI has a significant influence on undergraduate students performing brownfield development tasks in legacy code bases. 

Moreover, these findings align with Xu et al.'s~\cite{xu2022theory} theory-driven framework for AI-assisted programming, which suggests that AI tools can reduce cognitive barriers in programming by supporting attention management and knowledge retrieval. Our observed 34.9\% reduction in task completion time supports their theoretical prediction that GenAI assistance allows programmers to offload certain cognitive processes, particularly in brownfield contexts where the cognitive load of understanding existing code is substantial.

%We posit that the performance improvements stem from two key mechanisms: (1) reduced implementation overhead and a corresponding (2) reduction of cognitive load. In the simplest terms, by automating routine coding tasks, Copilot allows programmers to concentrate on solution refinement rather than low-level implementation details. Through the lens of cognitive load theory~\cite{sweller2011cognitive}, Copilot reduces \textit{extraneous cognitive load}---the mental effort expended on non-essential aspects of the task such as syntax recall and boilerplate code generation. This reduction frees up working memory for \textit{intrinsic cognitive load} (the inherent complexity of the programming task itself) and \textit{germane cognitive load} (the mental effort dedicated to learning and problem solving).

However, as noted by some of our participants, it appears that the GenAI assistant, rather than the human programmer, may actually be hefting much of the cognitive load of programming. Through the lens of cognitive load theory~\cite{sweller2011cognitive}, Copilot reduces \textit{extraneous cognitive load}---the mental effort expended on non-essential aspects of the task such as syntax recall and boilerplate code generation. This frees up working memory for \textit{intrinsic cognitive load} (the inherent complexity of the programming task itself) and \textit{germane cognitive load} (the mental effort dedicated to learning and problem-solving). However, when programming with GenAI, students may become cognitively disengaged with the programming process, accepting GenAI suggestions without reflecting on how they work or address the problem at hand. This robs students of crucial opportunities to learn and grow as programmers. 
% These findings support our first hypothesis (H1) that students would complete programming tasks more efficiently with Copilot assistance. The results also validate H2, predicting improved solution accuracy with AI assistance. This suggests that in educational contexts, AI tools like Copilot may serve not just as productivity enhancers but as scaffolding that allows students to focus on higher-level aspects of programming tasks.

\subsection{RQ2: Copilot's Impact on Programming Processes}

Copilot also profoundly impacted participants' programming workflows and interaction patterns. Our behavioral analysis revealed a restructuring of how students allocate their time and attention during programming tasks. The 10.63\% reduction in manual code writing and 11.6\% decrease in web search time indicate that Copilot serves dual functions as both a code generator and an information resource. This finding aligns with O'Brien et al.'s \cite{o2025scientists} observation that professionals often use GenAI tools as "information retrieval mechanisms" for programming support in their research. Furthermore, our finding that Copilot reduced manual coding comports with Shah et. al.'s ~\cite{shah2025students} observation that students perform ``one-shot prompting'' to have Copilot implement entire features at once. 

In addition, our Markov analysis identified a transition from a \textit{read~→~understand~→~implement} workflow without Copilot to a \textit{prompt~→~read response~→~implement} pattern with Copilot. This transition reflects what Liang et al.~\cite{Liang2023UnderstandingTU} call the "interactive steering" paradigm of GenAI programming assistance, where the programmer's role shifts toward guiding and evaluating GenAI-generated solutions rather than implementing solutions from scratch.

%The GenAI-mediated coding cycle (prompt writing → response viewing → code implementation) identified by our Markov analysis represents a fundamental shift in how students approach programming tasks. This workflow pattern replaces the traditional cycle of reading documentation, planning implementation, and manually writing code with a more iterative, GenAI-assisted approach.

%This suggests that GenAI assistance may encourage more empirical, outcomes-focused development strategies rather than implementation-focused approaches. As noted by S10, Copilot eliminates the "jumping back and forth between writing the code and searching," allowing for more continuous engagement with the programming task itself.

%These behavioral shifts raise important questions about how to update student learning outcomes in programming education to align with the skills needed to engage in the new GenAI-mediated coding cycle. Indeed, the traditional learning outcomes of programming assignments need to be updated to include the emerging metacognitive and critical thinking skills required to assess, refine, and iteratively integrate GenAI outputs into programming solutions. These skills seem especially important in the brownfield development contexts considered in this study, where code integration is tightly coupled with code base comprehension, and where  code solutions need to maintain consistency with the coding conventions and best practices of the existing code base.

\subsection{RQ3: Differences in how Copilot is Appropriated}
Our analysis of higher and lower performers revealed intriguing differences in how students strategically integrated Copilot into their workflow. Top performers were more selective in their use of AI-generated code, preferring granular inline suggestions over adoption of whole code blocks (83.9\% less pasting of Copilot code). This suggests that more discerning, critical engagement with AI suggestions may lead to better programming outcomes.

This finding aligns with research on metacognitive skills in programming. Loksa et al.~\cite{loksa2016programming} found that programmers with higher metacognitive awareness---the ability to plan, monitor, and evaluate their own cognitive processes---tend to make more strategic decisions during problem-solving. By critically evaluating GenAI suggestions rather than accepting them wholesale, our high performers appear to exhibit these metacognitive capabilities when interacting with Copilot. This selective integration approach reflects what Collins et al.\cite{collins1989cognitive} describe as "reflection" in their cognitive apprenticeship model, in which learners compare their own problem-solving processes against those of experts (or in this case, GenAI systems). Prather et al.~\cite{prather2024widening} observed a similar phenomenon in their study of novice programmers using GenAI tools, finding that ``students who already possessed strong metacognitive skills experienced enhanced performance" with GenAI assistance. This growing body of evidence underscores the importance of  metacognitive capabilities---specifically, an ability to maintain agency and critical judgment when presented with external solutions---as essential to the effective use of GenAI for programming.

% Qualitative data revealed a tension between productivity and comprehension that merits further investigation. While students like S03 found that Copilot made it "easier to find out where [I'm] supposed to be editing," others like S11 noted that Copilot "didn't help me understand necessarily" because "it just kind of gave me the answer." This tension highlights a potential trade-off between immediate task completion and deeper learning outcomes.

% S05's observation that students might be "better off without Copilot" when trying to understand a codebase raises important considerations for when and how to incorporate AI assistance in educational settings. This finding challenges our third hypothesis (H3) that students would demonstrate greater understanding of the legacy code base when working with Copilot. Instead, the relationship between AI assistance and code comprehension appears more nuanced, potentially varying based on individual learning strategies and the specific nature of programming tasks.

\subsection{Implications for Computing Education}
These findings have at least two key implications for computing education in the era of GenAI coding assistants:
\begin{enumerate}
    \item \textbf{Develop new learning outcomes and pedagogy for GenAI-assisted programming.} Given the new GenAI-assisted programming workflow (\textit{prompt~→~read response~→~implement}) reinforced by our study findings, computing educators need to rethink their pedagogical approach to teaching programming, along with the learning outcomes of that approach. While the ability to analyze problems and design and implement solutions will remain important learning outcomes, studies like ours suggest the need to define additional learning outcomes relevant to successful GenAI collaboration. New learning outcomes should target the prompt engineering, comprehension, metacognitive and critical thinking skills required to formulate programming questions and understand, assess, refine, and iteratively integrate GenAI outputs into programming solutions. These skills seem especially important in the brownfield development contexts considered in this study, where code integration is tightly coupled with code base comprehension, and where  code solutions need to maintain consistency with the coding conventions and best practices of the existing code base. Computing educators can then develop and evaluate new pedagogical methods to explicitly address these new learning outcomes.   
    \item \textbf{Balance assistance and learning:} The tension between productivity and learning highlights the need for carefully designed educational activities that encourage students to leverage GenAI assistance while still actively engaging in their own learning process. It is critical for computing educators to find ways to keep students cognitively and metacognitively engaged in GenAI-assisted programming, so that they think through GenAI suggestions and reflect on their relevance and correctness, rather than merely accepting GenAI solutions without critical thought or reflection. To that end, computing educators could consider introducing GenAI programming activities in which such critical thought and reflection are explicitly modeled and scaffolded. 
    % \item \textbf{Brownfield Pedagogy.} Traditional approaches to teaching legacy code development often focus on performing code comprehension before code modification. Our findings suggest that with GenAI assistance, this sequence might be reversed, with students leveraging GenAI to make  modifications first, potentially followed by deeper analysis of the existing codebase.
\end{enumerate}

These implications suggest that rather than simply restricting or embracing AI coding assistants, educators should develop nuanced approaches that strategically incorporate GenAI tools to facilitate clear learning outcomes while mitigating their potential drawbacks.

\section{Threats to Validity}
\label{sections-s6-threats}
\subsection{Internal Threats to Validity}

One threat to the internal validity of this study is that not all participants used the same version of the Visual Studio Code (VSC) GitHub Copilot extension. We failed to disable auto-updating of the Copilot extension, which evolved during the data collection period (June-November, 2024). A particularly significant change during this period was the introduction of "Copilot Edits", a feature that allows users to directly insert code from Copilot chat into their workspace. This feature, which became available in November 2024, created an inconsistency in the Copilot functionality accessible by participants; only one participant had access to Copilot Edits, and they used it during 40\% of their code writing time. While Copilot Edits did not fundamentally change the functionality of Copilot, it created a new way for one of the 10 participants to access Copilot and therefore could have influenced that participant's task success and programming process in the \textsc{Copilot} treatment.

A second threat to internal validity relates to the study's within-subjects design. By its very nature, a within-subjects design runs the risk of introducing a learning effect: Participants cannot unlearn what they learned in the treatment they experience first. A common approach to mitigating this learning effect is to counterbalance treatment order. For strategic reasons, we chose not to do that; we wanted to gauge participants' initial programming performance free from the influence of GenAI. By not counterbalancing treatment order, the learning effect of the No Copilot treatment went unmitigated. Along these same lines, a within-subjects study requires the tasks in each treatment to be as isomorphic as possible, to reduce the possibility that task disparities could influence results. We went to great lengths to equalize the difficulty and scope of tasks used in this study (see Section 3.3.2). In addition, we counterbalanced task order across participants to mitigate any task differences.

A third threat to internal validity relates to the limitations of our video data relative to our behavioral coding scheme. Some categorical definitions relied on visual evidence from the video regarding the location of the cursor; categorical shifts were defined by cursor movements to new windows on the screen. While it is reasonable to assume that participants' attention shifted with their moving cursor, this is not necessarily the case; eye tracking technology would have provided a more reliable indication of participants' switches to different behaviors. 

\subsection{External Threats to Validity}
Because of the time-consuming nature of the behavioral video coding process performed in this study, we elected to limit the number of participants to 10. While this is a small sample size for an experimental study, the study's within-subjects design and use of non-parametric statistical methods proved sufficient to obtain the statistical power necessary to detect significant differences between the study conditions. Nonetheless, the study's small sample size, drawn from the student population of a single university, suggests that extreme caution should be exercised in any attempt to generalize the results beyond the population we studied. 

Our study's generalizability is further limited by the technologies it used. The tasks focused on web programming (HTML/CSS/JavaScript), a single GenAI tool (GitHub Copilot), and limited development tasks. Our findings may not generalize  to  other programming languages, tasks, and GenAI tools.

Additionally, our participant sample was restricted to students who had demonstrated high academic performance in web development coursework, as evidenced by their grades (A or A-). This selective demographic profile means our findings may not be representative of the broader student population, with varying levels of academic achievement or programming aptitude.
%A second threat to the external validity of this study is our choice to focus on a particular GenAI tool (Copilot). While Copilot is widely used and tightly integrated into the VSC environment used by participants, it has specific features, such as auto-completion and in-line code querying, that may go beyond the type of GenAI support used by many computing students who rely exclusively on chat. Thus, one must be careful to generalize these results beyond the Copilot extension used in this study.

\section{Summary and Future Work}
\label{sections-s7-conclusions}
In this paper, we have presented an experimental comparison of student programmers' outcomes and processes when working on brownfield programming tasks with and without the help of a GenAI programming assistant. We can identify four key takeaways from our results:

\begin{itemize}
    \item In brownfield programming tasks, students' use of GenAI promotes significant speed and productivity gains as compared to not using GenAI.
    \item GenAI significantly changes students' programming processes, shifting them from the traditional\textit{ read~→~understand~→~implement} workflow to a  \textit{prompt~→~view response~→~implement} workflow. 
    \item Higher-performing students tend to be more selective than lower-performing students in their use of GenAI, spending higher percentages of their implementation time integrating autocomplete suggestions into their code and performing manual edits.
    \item Some students lament that they use GenAI without really understanding  the code it generates or how that code fits into the legacy code base they are editing, suggesting that the programming efficiencies afforded by GenAI may be offset by a diminished understanding of how and why GenAI-generated code actually works. 
\end{itemize}

% These findings suggest several directions for future computing education research. First, given the productivity-learning tradeoff that GenAI may promote, at least for some students, future research should investigate ways to  disconnect between GenA. First,    

This study reveals several directions for future computing education research. Longitudinal studies are needed to examine how prolonged Copilot usage affects students' programming skill development and code comprehension abilities over time. Future work should employ larger and more diverse participant samples to increase generalizability, while implementing methodological improvements such as eye-tracking, and think-aloud protocols specifically targeting AI interaction decisions with more probing into why students may or may not adopt certain AI suggestions. Comparative studies across different GenAI tools could identify which features most effectively support brownfield programming tasks. These research directions would address the fundamental tension between the productivity benefits of GenAI and the deeper understanding required for effective legacy code development in professional software development.

\section{Acknowledgement}
\label{sections-s8-acknowledgement}
% Acknowledgments are placed before the references. Add information about grants, awards, or other types of funding that you have received to support your research. This information must be anonymized in the version of the paper you submit for review.  
% Acknowledgments omitted for anonymous review.
We would like to express our sincere gratitude to Dr. Christopher Sanchez for his invaluable assistance during the study design period. His expertise and guidance were instrumental in shaping the methodological framework of this research. We also thank all participants who volunteered their time to contribute to this study. Finally, we acknowledge the support of Oregon State University in providing the necessary resources and infrastructure for conducting this research.

% \section{ACKNOWLEDGMENTS}
% \label{sections-s8-acknowledgement}
% \input{sections/s8-acknowledgement}
%%
%% The acknowledgments section is defined using the "acks" environment
%% (and NOT an unnumbered section). This ensures the proper
%% identification of the section in the article metadata, and the
%% consistent spelling of the heading.
% \begin{acks}
% To Robert, for the bagels and explaining CMYK and color spaces.
% \end{acks}

%%
%% The next two lines define the bibliography style to be used, and
%% the bibliography file.
% \bibliographystyle{unsrt}
\bibliographystyle{ACM-Reference-Format}
\bibliography{main}

\end{document}